\documentclass[final,1p,times]{elsarticle}

\usepackage{graphics}

\usepackage{amssymb}
\usepackage{amsmath}
\usepackage{amsthm}

\usepackage{bm}

\def\be{\begin{eqnarray}}
\def\ee{\end{eqnarray}}
\def\bc{\begin{center}}
\def\ec{\end{center}}
\def\trm{\textrm}

\def\LamFOF{\Lambda \trm{(1405)}}
\def\LamStar{\Lambda ^{\ast}}
\def\LamN{\Lambda N}
\def\SigzN{\Sigma ^{0} N}

\def\Lamp{\Lambda p}
\def\Sigzp{\Sigma ^{0} p}
\def\Sigpn{\Sigma ^{+} n}
\def\Lamn{\Lambda n}
\def\Sigzn{\Sigma ^{0} n}
\def\Sigmp{\Sigma ^{-} p}
\def\KbarN{\bar{K} N}
\def\piS{\pi \Sigma}

\def\etL{\eta \Lambda}

\journal{Nuclear Physics A}

\begin{document}

\begin{frontmatter}



\title{$\Lambda \text{(1405)} N \to Y N$ transition in nuclear medium for 
non-mesonic absorption of $\bar{K}$ in nucleus} 


\author[label1]{T. Sekihara}
\author[label2]{D. Jido}
\author[label2]{Y. Kanada-En'yo}

\address[label1]{Department of Physics, Kyoto University, Kyoto 606-8502, Japan}
\address[label2]{Yukawa Institute for Theoretical Physics, 
  Kyoto University, Kyoto 606-8502, Japan}


\begin{abstract}
Non-mesonic transition of $\Lambda \text{(1405)} N \to YN$ is investigated
as one of the essential processes for the non-mesonic absorption of $\bar{K}$ 
in nuclei. Using one-meson exchange model in the calculation of the 
transition, we find that the non-mesonic transition ratio 
$\Gamma_{\Lambda N} / \Gamma_{\Sigma N}$ depends strongly on the ratio of the 
$\Lambda \text{(1405)}$ ($\LamStar$) couplings to $\bar{K} N$ and 
$\pi \Sigma$. Especially a larger $\LamStar$-$\bar{K} N$ coupling leads to 
enhancement of the transition to $\Lambda N$. Using the chiral unitary 
model for the description of the $\LamStar$, we obtain 
$\Gamma_{\Lambda N}/ \Gamma_{\Sigma ^{0} N} \approx 1.2$ 
which is almost independent of the nucleon density, and find 
the total non-mesonic decay width of the $\LamStar$ 
in uniform nuclear matter to be $22 \, \text{MeV}$ at the normal density. 
\end{abstract}


\end{frontmatter}



\section{Introduction}
\label{sec:Introduction}

The study of the in-medium properties of anti-kaon ($\bar{K}$) has attracted 
continuous attention~\cite{kaon}. 
$\bar{K}$ has been theoretically expected to 
be bound by nuclear systems due to attractive strong interaction between 
$\bar{K}$ and nucleon $N$~\cite{theories}. 
These $\bar{K}$-nuclear bound systems assisted mainly by the strong 
interaction, which are so-called kaonic nuclei, have at least two important 
aspects; the kaonic nuclei are strongly interacting exotic many-body systems, 
and they provide favorable sysmets for the studies of the $\bar{K}$ properties 
at finite density. However, in spite of experimental efforts to search for 
the $\bar{K}$-nuclear bound systems~\cite{experiments}, 
there are no clear evidences observed yet. 

In order to understand the $\bar{K}$-nucleus systems, it is 
important to investigate theoretically the decay mechanism of 
the kaonic nuclei, or the absorption process of the $\bar{K}$ into 
nuclei. Especially the decay (absorption) ratios of various modes 
and explicit numbers of the decay (absorption) widths are 
key quantities for the $\bar{K}$-nucleus systems. The decay of the 
kaonic nuclei in strong interactions can be categorized into two processes; 
one is the mesonic process such as $\KbarN \to \pi Y$, and the other is the 
non-mesonic process such as $\KbarN N \to Y N$, where $Y$ denotes hyperon 
($\Lambda$ or $\Sigma$). The non-mesonic decays have advantage in experimental 
observations, since signals from kaonic nuclei are readily distinguished from 
backgrounds and no extra mesons do not have to be detected. Therefore, 
systematic studies of the non-mesonic decay of the kaonic nuclei are 
desirable. Especially the ratios of the decay widths are interesting, since 
they will be insensitive to details of the production mechanism. 

The absorption of the $\bar{K}$ in nuclei may take place dominantly 
through the $\Lambda (1405)$ ($\LamStar$) resonance, owning to the presence 
of the $\LamStar$ just below the $\KbarN$ threshold. 
Namely the $\LamStar$ can be a doorway of the $\bar{K}$ absorptions 
in nuclei. The $\LamStar$ doorway picture is more probable, 
in case that the $\LamStar$ is a quasi-bound state of 
$\KbarN$~\cite{Dalitz:1967fp,Hyodo:2008xr}, 
which has large $\KbarN$ components as almost real particles. 
Therefore the strong $\KbarN$ correlations are expected to be 
responsible for the $\LamStar$-induced decays of the $\bar{K}$ in nuclei. 
Motivated by the $\LamStar$ doorway picture, we study the non-mesonic 
transition of $\LamStar N \to YN$ ($\LamStar p \to \Lamp, \, \Sigzp , \, 
\Sigpn$ and $\LamStar n \to \Lamn , \, \Sigzn , \, \Sigmp$)
in uniform nuclear matter with a one-meson exchange approach.

\section{$\bm{\LamFOF}$ doorway process for the non-mesonic decay in kaonic 
nuclei}
\label{sec:Doorway}

Now let us define the transition rate of the $\LamStar N \to Y N$ process by 
the transition probability divided by time $\mathcal{T}$ as, 
\be
\gamma _{YN}   \equiv 
\frac{1}{\mathcal{T}} \frac{1}{\mathcal{V}^{2}} 
\frac{1}{4} \sum _{\trm{spin}} 
\int d\Phi_{2}
|S - 1|^{2} 
= 
\frac{1}{\mathcal{V}} 
\frac{1}{4} \sum _{\trm{spin}} 
\int d\Phi_{2}
|T_{YN}|^{2} 
(2 \pi)^{4} \delta ^{4} (P_{\LamStar} + P_{\trm{in}} - P_{N} - P_{Y}) , 
 \label{eq:transition-rate}
\ee
with the $S$-matrix $S$ for the transition process given 
by the transition amplitudes $T_{YN}$ as $S=1-i
 (2 \pi)^{4} \delta ^{4} (P_{\LamStar} + P_{\trm{in}} - P_{N} - P_{Y}) 
 T_{YN}$. $\mathcal{V}$ is volume of the system, and $d\Phi _{2}$ is the 
phase-space for the final state.  The transition rate depends on the 
center-of-mass energy $E_{\text{c.m.}}$, equivalently the initial nucleon 
momentum. Summing up the $\LamStar N \to Y N$ transition rate in terms 
of the initial nucleon states, we can estimate the non-mesonic 
decay width of the $\LamStar$ in nuclear medium as a function
of the nucleon density. The factor $1/\mathcal{V}$ in the last form of 
Eq.~(\ref{eq:transition-rate}) is responsible for the fact that only one $N$ 
exists in the initial state in the volume $\mathcal{V}$. 

\begin{figure}[!t]
  \centering
  \begin{tabular}{@{\extracolsep{\fill}}cccc}
    \includegraphics[scale=0.12]{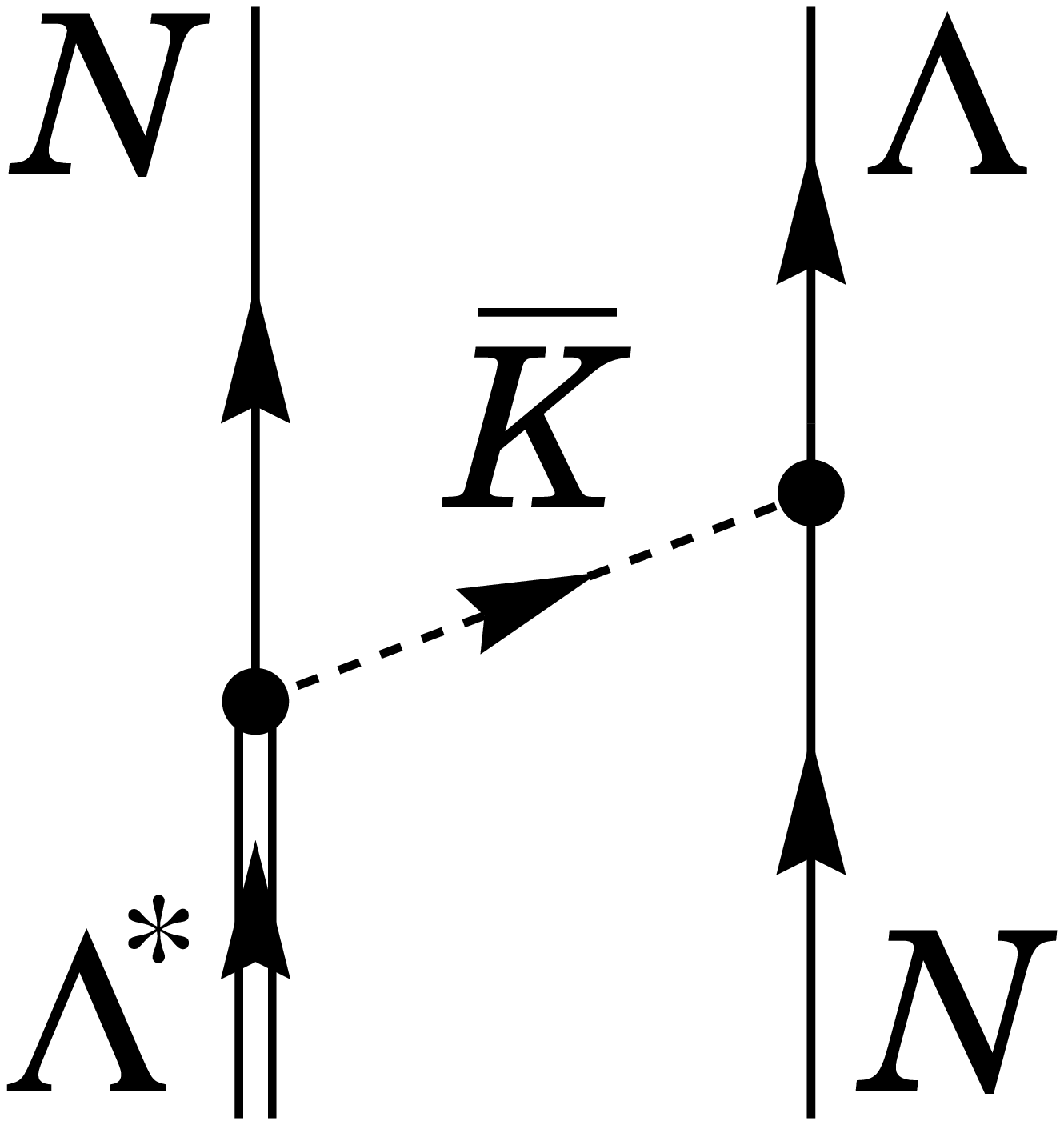}
    &
    \includegraphics[scale=0.12]{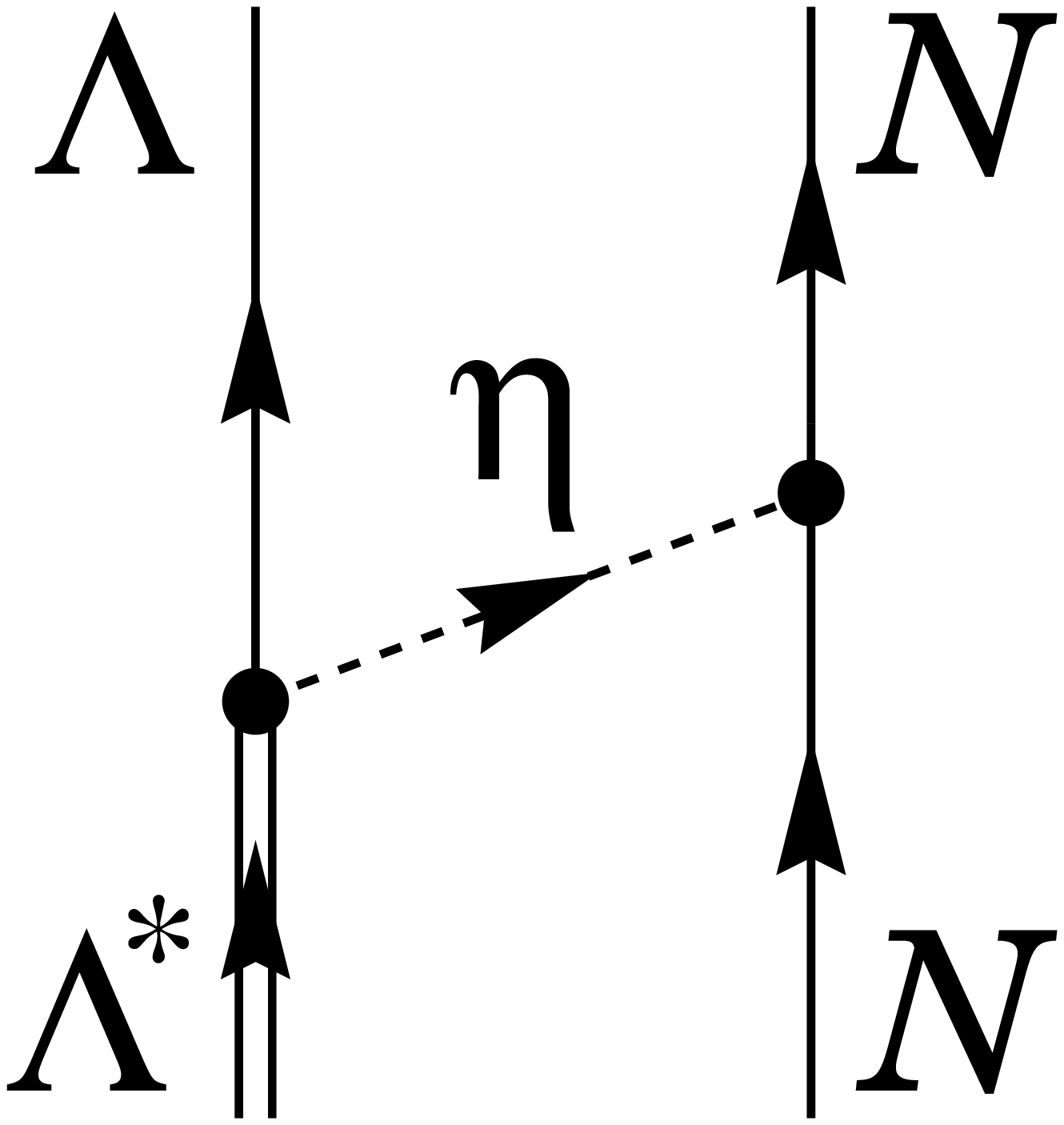} 
    &
    \includegraphics[scale=0.12]{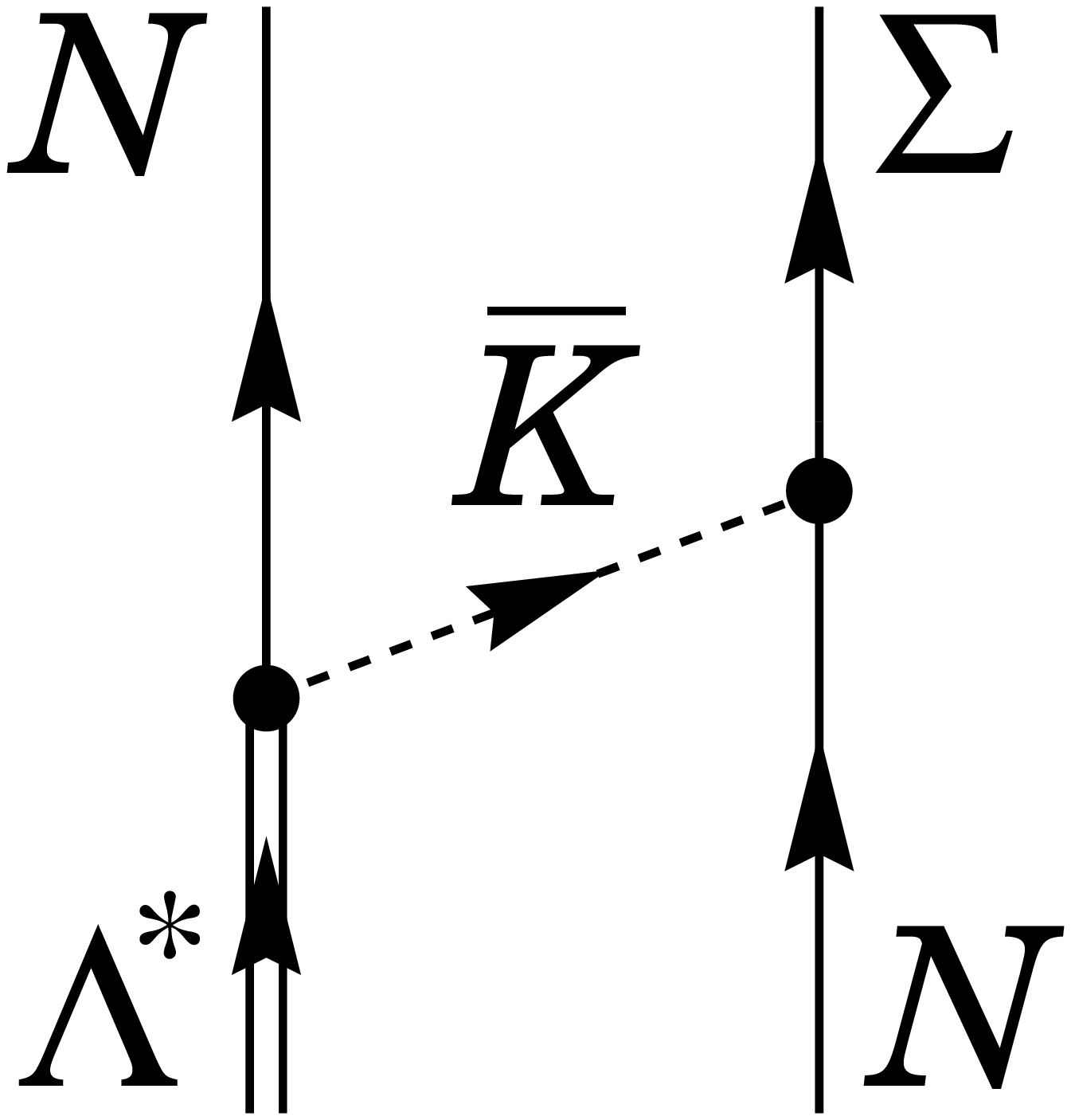}
    &
    \includegraphics[scale=0.12]{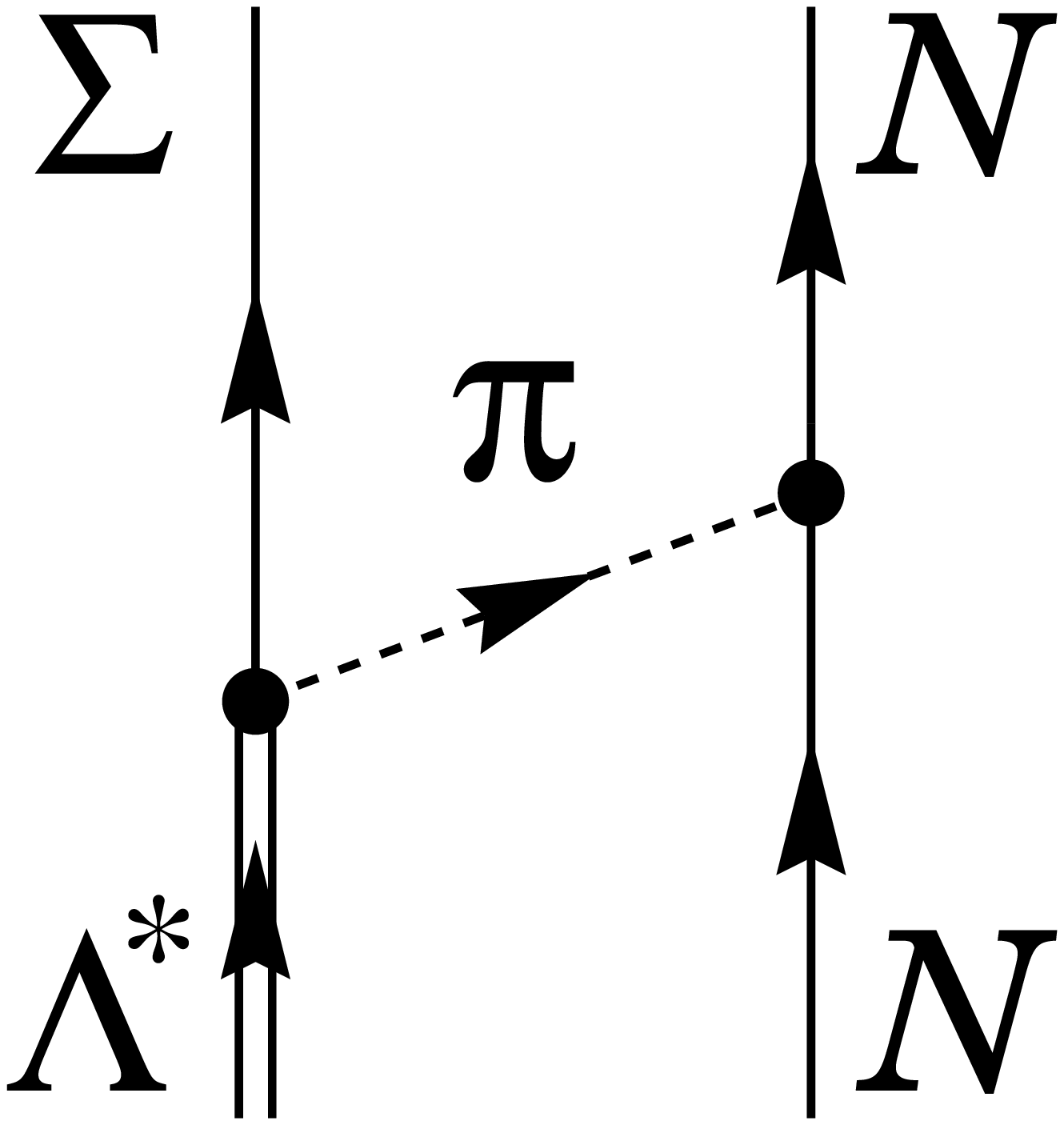} \\ 
    $\LamN 1$ & $\LamN 2$ &
    $\Sigma N 1$ & $\Sigma N 2$ 
  \end{tabular}
  \caption{Feynman diagrams for the $\LamStar N \to YN$
    transition in the one-meson exchange model. 
    The left two diagrams are for the $\Lambda N$ final state
    and the right two are for the $\Sigma N$ state. }
  \label{fig:non-M}
\end{figure}

We evaluate the transition amplitudes $T_{YN}$ with one-meson exchange 
diagrams shown in Fig.~\ref{fig:non-M}. The each diagram is composed of 
three parts: the $s$-wave $\LamStar M B$ coupling $G_{MB}$, the meson 
propagator including short-range correlation 
$\tilde{\Pi}(q^{2})$~\cite{Oset:1979bi,Jido:2001am}, 
and the $p$-wave $MBB$ 
coupling $V_{MBB}$, which is determined by the flavor $\text{SU}(3)$ symmetry. 
The explicit form is given in Ref.~\cite{Sekihara:2009yk}. 

\begin{figure}[!t]
  \centering
  \includegraphics[scale=0.22,angle=-90]{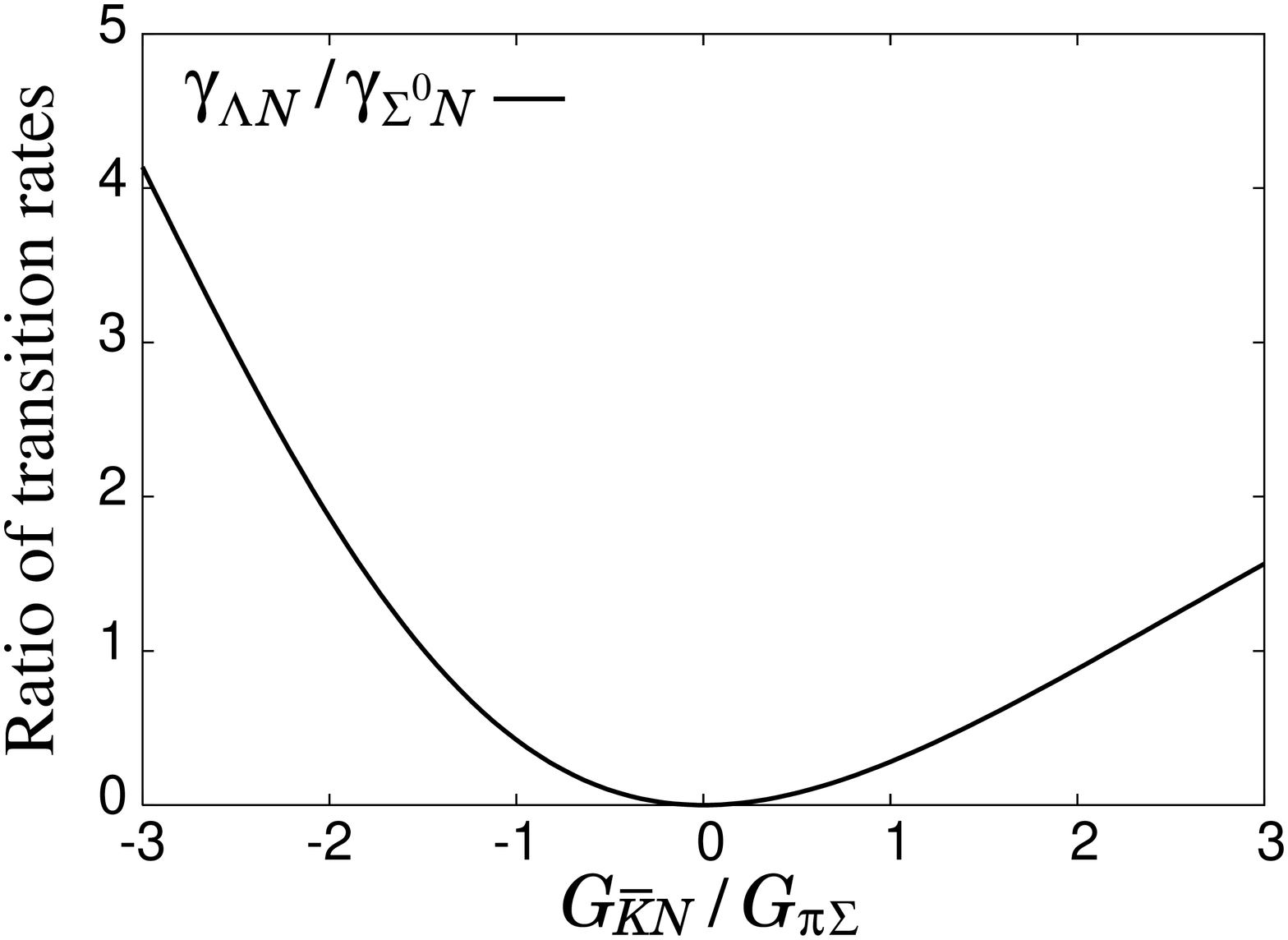}
  \caption{Ratio of the transition rates of $\LamStar N$ to 
    $\Lambda N$ and $\Sigma^{0} N$, $\gamma _{\LamN} / 
    \gamma _{\SigzN}$, as a function of $G_{\KbarN}/G_{\piS}$.
    We fix the initial nucleon momentum $ p_{\trm{in}} =0$ and 
    the $\LamStar$ mass $M_{\LamStar}=1420 \, \trm{MeV}$
    and assume  $G_{\etL}=0$.}
  \label{fig:GratiovsCratio}
\end{figure}

Among the three parts for the transition, only the $s$-wave $\LamStar MB$ 
coupling constants $G_{MB}$ are the model parameters, which are determined 
by the properties of the $\LamStar$. Therefore, in order to study the 
non-mesonic decay pattern in the $\LamStar$ doorway, let us discuss the 
$G_{MB}$ coupling dependence of the ratio of the transition 
rates, $\gamma _{\LamN} / \gamma _{\SigzN}$. With the conditions that $G_{\etL}=0$, 
the initial nucleon momentum $ p_{\trm{in}} =0$, and the $\LamStar$ mass 
$M_{\LamStar}=1420 \, \trm{MeV}$, we show the numerical result of 
the ratio of the transition rates, $\gamma _{\LamN} / \gamma _{\SigzN}$ 
as a function of $G_{\KbarN}/G_{\piS}$ in Fig.~\ref{fig:GratiovsCratio}, in 
which we assume $G_{\KbarN}/G_{\piS}$ to be a real number. As seen in the figure, 
the ratio $\gamma _{\LamN} / \gamma _{\SigzN}$  has 
strong dependence on the coupling ratio $G_{\KbarN}/G_{\piS}$. 
This is because the transition to $\Lambda N$ is governed by
the $\bar K$ exchange and the transition to $\Sigma N$ 
is dominated by the $\pi$ exchange due to the coupling $V_{MBB}$ strength. 
This result suggests that larger $\LamStar \KbarN$ 
coupling leads to enhancement of the decay ratio to $\LamN$ 
in the kaonic nuclei, although the $\LamStar$ 
in vacuum cannot decay into final states including $\Lambda$. 

Now we calculate the non-mesonic decay of the $\LamStar$ in uniform 
nuclear matter induced by the $\LamStar N \to YN$ transition, 
under the free Fermi gas approximation for nuclear matter. 
We evaluate the non-mesonic decay width $\Gamma _{YN}$ 
by summing up the transition rate $\gamma _{YN}$ for the nucleons: 
\be
\Gamma _{YN}  \equiv \sum _{i=1}^{A_{N}} 
\gamma _{YN} (k_{i}) 
= 
\int _{0}^{k_{\text{F}N}} d k \frac{k^{2}}{\pi ^{2}} 
\, \mathcal{V} 
\, \gamma _{YN} (k) , 
\ee
where $A_{N} = k_{\text{F}N}^{3} / (3 \pi ^{2}) \times \mathcal{V}$ 
($N = p$ or $n$) is the numbers of the protons or neutrons in $\mathcal V$. 

Furthermore we fix the ratios of the $\LamStar MB$ coupling constants from 
the $s$-wave $\KbarN (I=0) \to MB (\text{particle-basis})$ scattering 
amplitudes $t_{MB}(W)$ around the $\LamStar$ resonance energy, as ratios of 
the $t_{MB}(W)$. Here $W$ is the 
center-of-mass energy chosen to be the $\LamStar$ resonance position on the 
real energy axis. For the description of the $\KbarN$ scattering amplitudes 
$t_{MB}(W)$, we use the chiral unitary approach (ChUA), which reproduces 
well the $s$-wave scattering amplitudes of $\KbarN \to M B$
in the coupled-channels method based on chiral dynamics with dynamically 
generated 
$\LamStar$~\cite{chiral}. 
We use the parameter set in Ref.~\cite{Sekihara:2008qk} and take 
$W=M_{\LamStar}=1420 \, \trm{MeV}$, which is obtained by chiral unitary approach.

\begin{figure}[!b]
  \centering
  \begin{tabular}{@{\extracolsep{\fill}}cc}
    \includegraphics[scale=0.22,angle=-90]{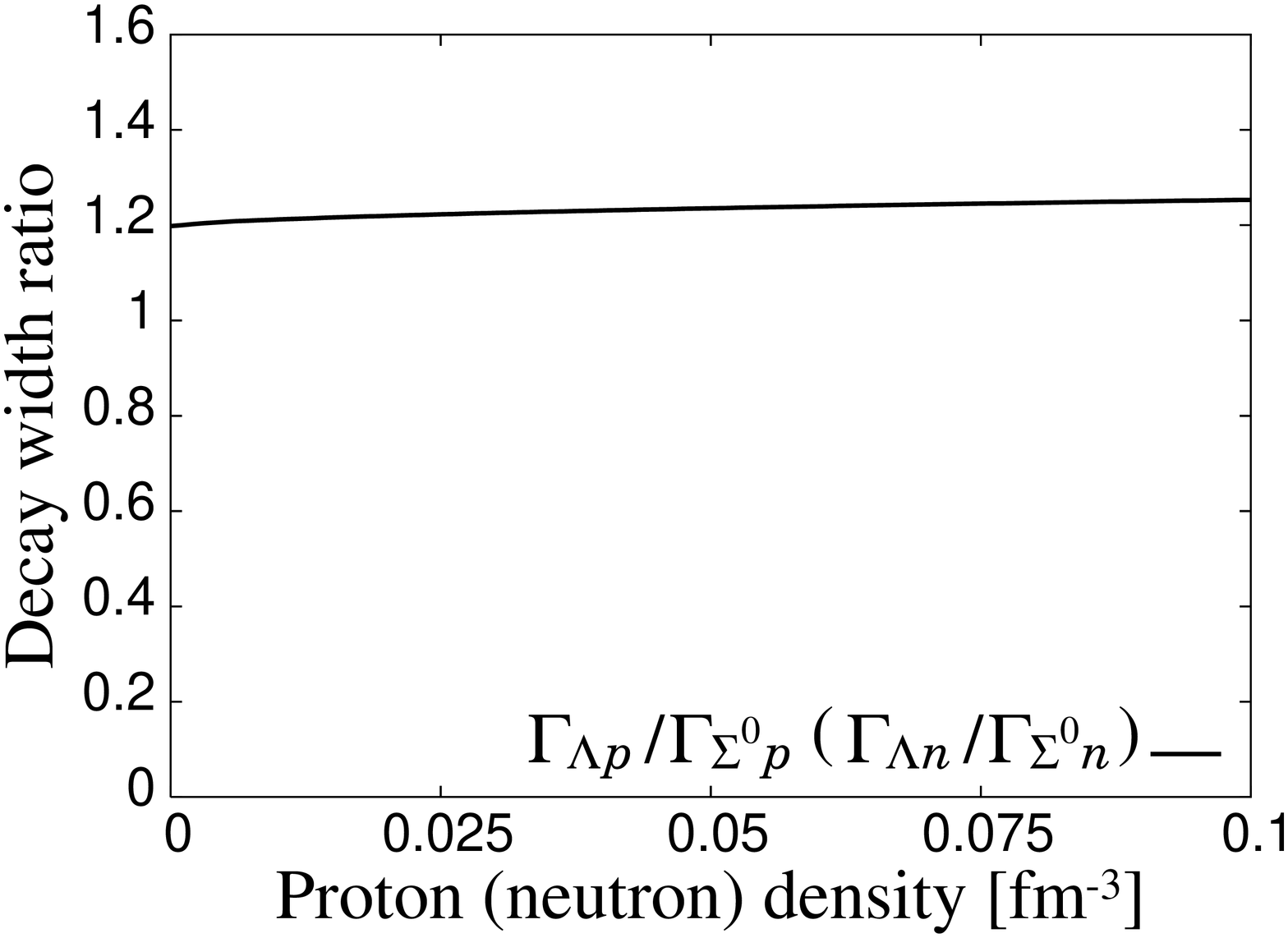} & 
    \includegraphics[scale=0.22,angle=-90]{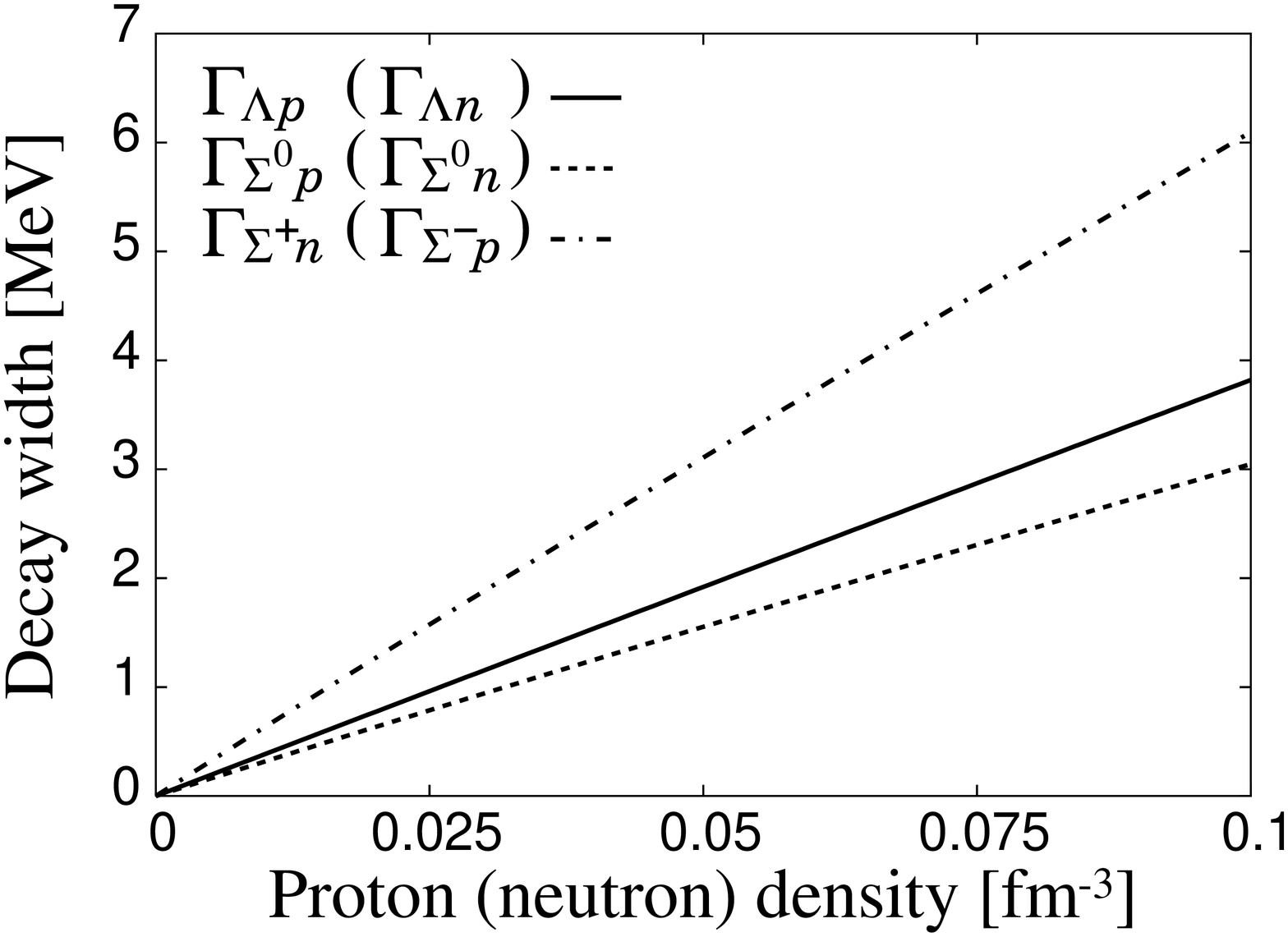} \\
    (a) & (b)
  \end{tabular}
  \caption{(a) Non-mesonic decay width ratio of $\LamStar$ in nuclei. 
    (b) Non-mesonic decay width of $\LamStar$ in nuclei. In both cases 
    the $\LamStar$ coupling constants are determined by the chiral unitary 
    approach. }
  \label{fig:GvsRho}
\end{figure}

Fixing the ratios of the $\LamStar$ coupling constants, $G_{MB}$, 
we plot the ratio of the non-mesonic decay widths of the 
in-medium $\bar{K}$ to $\Lambda N$ and $\Sigma^{0}N$ as a function of the 
proton (or neutron) density in Fig.~\ref{fig:GvsRho}(a). 
This figure shows that the ratio of the non-mesonic decay widths 
$\Gamma _{\LamN} / \Gamma _{\SigzN}$ is around $1.2$ 
almost independently of the nucleon density. 
The density independence of the ratio of the decay widths is 
caused by sufficiently large phase-space in the final states. 

We also obtain the absolute values of the non-mesonic decay
widths of the $\LamStar$ in nuclear matter. We use the couping 
constants of the $\LamStar$ to $\bar{K} N$, $\pi \Sigma$ and 
$\eta \Lambda$ obtained by the chiral unitary approach, which 
gives the in-vacuum mesonic decay as $\Gamma_{\LamStar \to \pi \Sigma}=40 \, 
\text{MeV}$ with $M_{\LamStar}=1420 \, \text{MeV}$ as observed in a $K^{-}$ 
initiated channel~\cite{Braun:1977wd,Jido:2009jf}. 
We show the non-mesonic decay widths in Fig.~\ref{fig:GvsRho}(b). 
The linear dependence of the decay widths is caused by the large 
phase-space in the final states. 
At the normal nuclear density ($\rho _{\trm{B}} = 0.17 \, \trm{fm}^{-3}$),
the total non-mesonic decay width is $22 \, \trm{MeV}$, 
which is almost half of the mesonic decay width of $\LamStar$ 
($\sim 40 \, \trm{MeV}$). 

\section{Conclusion}
\label{sec:Conclusion}

We have investigated the non-mesonic transition $\LamStar Y 
\to Y N$ in nuclear medium as one of the essential processes for the 
non-mesonic absorption of $\bar{K}$ in nucleus. Calculating the 
$\LamStar N \to YN$ transition rates in the 
one-meson exchange processes, we have found that the ratio of the 
$\LamStar N$ transition rates to $\Lambda N$ and $\Sigma N$ strongly 
depends on the ratio of the $\LamStar$ couplings, $G_{\bar KN}/G_{\pi\Sigma}$. 
Especially, larger $\LamStar$ couplings to $\KbarN$
lead to enhancement of the non-mesonic $\LamStar$ decay with 
$\Lambda N$ emission. Furthermore, describing the $\LamStar$ properties by 
the chiral unitary approach, we have obtained the ratio of the non-mesonic 
decay widths to $\Lambda N$ and $\Sigma ^{0} N$ as 
$\Gamma _{\LamN} / \Gamma _{\SigzN} = 1.2$ almost independently 
of the nucleon density. We have also estimated that the total 
non-mesonic decay width is $22 \, \trm{MeV}$ at the saturation density. 
This study can be extend to the calculation of the absorption
width of the $\bar K$ in nuclear medium~\cite{preparation}. 


\end{document}